\documentclass{elsart}
\usepackage{graphicx}

\begin{document}
\begin{frontmatter}
\title{Superconductivity Induced by Negative Centers }

\author[label1]{G. Litak\thanksref{E-mail}}
\author[label2]{T. Paiva},
\author[label3]{R.T. Scalettar},
\author[label4]{C. Huscroft}, 
\author[label2]{and R.R.  dos Santos}

\address[label1]{Department of Mechanics, Technical University of
Lublin, Nadbystrzycka 36, PL--20-618  Lublin, Poland}

\address[label2]{Instituto de Fisica, Universidade Federal do Rio de
Janeiro, Cx.P. 68 528, 21945-970 Rio de Janeiro RJ, Brazil}

\address[label3]{Physics Department, University of California, Davis, CA 
95616}

\address[label4]{Hewlett-Packard Company, 8000 Foothills Boulevard,
Roseville, CA 95747}

\thanks[E-mail]{Fax: +48-815250808; E-mail:
litak@archimedes.pol.lublin.pl}

\begin{abstract}
We study the effect of random dilution of attraction centers on electron pairing
in a generalization of the negative $U$ Hubbard model 
where the interaction $U_i$, defined on the
two dimensional lattice, can be  $-|U|$
and 0 with probability $c$ and $1-c$ depending on the lattice site $i$. 
Using the determinant quantum Monte Carlo approach,
we find the critical concentration $c_0$
of negative centers which leads to superconductivity, and show how
the evolution of the local density on the two types of sites lends
insight into the formation of global phase coherence.   

\end{abstract}

\end{frontmatter}

\section{Introduction}

The behavior of High Temperature Superconductors (HTS) is
characterized by a strong dependence on doping.
For example, changing the Sr concentration in La$_{2-x}$Sr$_x$CuO$_4$,
or the O concentration in YBa$_2$Cu$_3$O$_{6+\delta}$
tunes these materials from antiferromagnetic insulators, to
spin glasses, and finally to superconductors.
The short coherence length and relatively low superfluid density
in the underdoped region emphasize the existence of phase fluctuations and 
enhance the competition  between  superconductivity and other
long range order phases \cite{Mic97,Eme95}. 
In such systems, where different ground states emerge as the
particle density is tuned, it is often important to explore the  possibility 
inhomogeneous mixtures of phases; for example, 'antiferromagnetic' 
and 'superconducting', at commensurate and incommensurate filling. 
Indeed, there has already been considerable study of whether the
doped holes are uniformly distributed or whether they
accumulate preferentially in two-dimensional
phase separated regions\cite{Lin90} or in one-dimensional 
stripes\cite{Scalapino01}.

There have been many direct studies of doping dependent transitions
in correlated electron Hamiltonians 
like the repulsive Hubbard model, which 
provide microscopic models of HTS\cite{Dagotto94,Scalapino01}.
These are numerically challenging since they attempt to
understand the origin of HTS at the same time as addressing
additional subtle issues like inhomogeneous phases.
A more simple, if somewhat less fundamental, way in which related questions can be 
addressed is to consider models, like the
$-U$ Hubbard Hamiltonian, which have an attractive interaction
`built in'.  This separates out the problem of how
superconducting pairs form in disordered, short coherence length systems
in which the particle density is fixed 
but the interactions are diluted,  from the complexities of
the origins of the attractive interaction.
The key underlying question is how superconductivity spreads through a system as the
local attraction between charge carriers is strengthened. 
Indeed, such centers of attraction played important role in the several models
of HTS\cite{Wil89,Mic90,Wil00,Tin92,Ale95,Ale99,Ran95,Ran98,Tri98}. 
Another possible experimental realization is the percolative, 
granular-like transition to superconductivity via hydrogen doping 
which has been reported in
Eu$_{1.5}$Ce$_{0.5}$NbSr$_{2}$Cu$_{2}$O$_{10}$\cite{Levi99},  
a ceramic with  the superconducting
critical temperature $T_c$= 29 K. 

In this paper, we shall adopt this approach and study a simple model 
which allows us to address the issue of superconducting transitions tuned
by changing the concentration of attraction centers.  
Specifically, we study a two-dimensional
negative $U$ Hubbard model which has an on-site attraction on
some fraction of its lattice sites and ask what fraction of sites 
is needed to be attractive for the system to superconduct.  
Recently, this issue has been addressed  by Litak and Gy\"{o}rffy 
\cite{Lit00} who  examined the problem of superconducting percolation
by  using Hartree-Fock-Gorkov decoupling as an approximation for 
interacting electrons and Coherent Potential Approximation (CPA) to treat 
randomly distributed centers.
They  have  found the critical concentration of negative centers
$c_0=n/2$, where $n$ denotes a band filling in the limit of large
$|U|$ interaction. For
a small concentration of centers $c < c_0$ the system becomes normal
because every site, where   $-|U|$ interaction is present, is doubly occupied.

The effect of dilution of the interactions in the attractive $U$ Hubbard
model at half-filling, $n=1$ can also be inferred from the 
Quantum Monte Carlo work of Ulmke {\it et al.}\cite{Ulm98}, who
investigated the suppression of antiferromagnetism by $U=0$ impurities in
the repulsive  Hubbard model ( $U > 0$) at $n=1$.  
The critical concentration $f_c$ of 
impurities which destroy the antiferromagnetism was
estimated as $f_c \approx 0.45$.
Since a particle-hole transformation maps the 
attractive and repulsive $U$ Hubbard models onto one another,
at half-filling, this work immediately identifies the
critical concentration $c_0$ for suppression of superconductivity
in the attractive Hubbard model, $c_0=1-f_c=0.55$.

Further analytic work on the disordered attractive
Hubbard model, for example based on numerical
solution of the Bogliubov-de Gennes equations, is contained in
\cite{Gho98,Tri01,Gyo91}.
However, there have to date been no Quantum Monte Carlo (QMC)
studies for the doped case where the density is
not precisely half-filled and particle-hole symmetry is broken.  
In this paper
we examine the problem of percolating superconductivity using QMC
in the negative $U$ Hamiltonian changing the concentration
of negative centers $c$ as well as band filling $n$\cite{Lit00}. 
The use of QMC enables us to go beyond the limitations of mean
field approximations
and  investigate the effect of charge and pairing potential  fluctuations
on an equal footing. 

The paper is organized as follows. After an introduction in Sec.~1
we present  the microscopic model and formulate the problem (Sec.~2). 
Section 3 defines the QMC
Monte Carlo method used in simulations and shows the key
numerical results: the prediction of superconductivity 
percolation and the critical concentration $c_0$.
The paper ends with conclusions.    

\section{The microscopic model}  
   
We consider the
random negative $U$ Hubbard Model \cite{Lit00} defined by the following
Hamiltonian:
\begin{equation}
\hat H= -\sum_{i j \sigma}t_{i j} c_{i \sigma}^{+} c_{j \sigma} +
\frac{1}{2} \sum_{i \sigma}  U_i  \,\, c_{i \sigma}^{+} c_{i \sigma} c_{i  
-\sigma}^{+}
c_{i -\sigma} - \mu\sum_{i \sigma} c_{i\sigma}^+c_{i\sigma}~ . \label{E1}
\nonumber
\end{equation}
Here $i$ and $j$ label sites on a two-dimensional
square lattice, $t_{ij}$ is the hopping integral
connecting only nearest neighbour $i$ and $j$'s, $\mu$ is the electronic
chemical potential, $c_{i\sigma}^+$,  $c_{i\sigma}$ create and
annihilate, respectively, electrons at the single site $i$ with spin
$\sigma$, and the coupling constant,
\begin{equation}
U_{i}= \left\{ \begin{array}{rl} -|U| & {\rm with~ probality}~ c
\\ 0  & {\rm with~ probability}~ 1-c
\end{array}~. \right. \label{E2}
\end{equation}

\begin{figure}[htb]
\centering
\vspace{0.45cm}
\resizebox{0.7\textwidth}{!}{%
\includegraphics{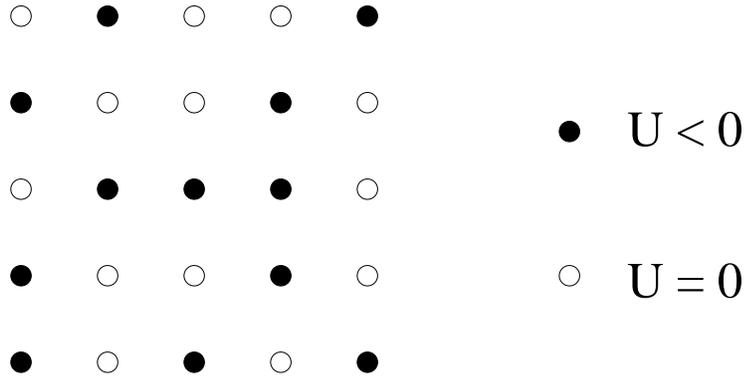}}
\vspace{1cm}
\caption{The two dimensional lattice with random electron interaction
$U= 0$ or $-|U|$.}
\end{figure}

We are interested in determining, away
from particle-hole symmetry, the critical concentration $c_0$ for the $U$ centers
such that for $c < c_0$ the configurationally averaged, superconducting
long range order parameter vanishes.
At the same time, we would like to obtain some insight into
the mechanism of the formation of long range order as one
moves to $c>c_0$.

Such questions can be addressed by computing expectation values
of operators like the Hamiltonian, the kinetic energy and double
occupation, and pair-pair correlation functions.
Such expectation values take the general form,
\begin{equation}
\langle \hat A \rangle = { {\rm Tr} \hat A {\rm e}^{-\beta \hat H} \over
{\rm
Tr}
{\rm e}^{-\beta \hat H} }~. \label{E3}
\end{equation}
and can be evaluated by a standard procedure \cite{Bla81,Hus97,Hus98,Sca99}
which involves writing a path integral expression
for the imaginary time evolution operator $e^{-\beta H}$ 
and performing a stochastic integration over a set
of auxiliary fields introduced to treat the
electron-electron interaction exactly.
This will be described in greater detail in the following section.
Typically, we can study system of a few hundred electrons,
far larger than the competing exact diagonalization approach,
but still small enough to require careful treatment
of finite size effects, as we shall discuss below.

We will focus especially on
the equal time pair correlation function, which is defined by
\begin{equation}
p_s(j-l)= < \Delta_l \Delta_j^+ >, ~~~{\rm where}~~~
\Delta_j^+=c_{j \uparrow}^+c_{j \downarrow}^+ \label{E4}
\end{equation}
Here $\Delta_j^+$ creates a pair of electrons at lattice site $i$.
The long range order in $p_s(l)$
can be probed by calculating its structure factor,
\begin{equation}
S_p= \frac{1}{N} \sum_{jl} p_s(j-l)~. \label{E5}
\end{equation}
A finite positive value $S_p > 0$, in the limit of large lattice size $L$,
signals superconductivity, otherwise the system is normal.
For our disordered system (Eq. \ref{E2}, Fig. 1) we have to
do a disorder average over many realizations of random
configurations of the attractive $U$ centers.

\section{Details of Quantum Monte Carlo Approach}

In order to describe 
the quantum Monte Carlo method \cite{Bla81} in
more detail,
let us first rewrite the Hubbard Hamiltonian in
a way which puts the interaction term in a particle-hole
symmetric form.
\begin{equation}
\hat H=-t \sum_{ ij\sigma}
c_{i\sigma}^{\dagger} c_{j\sigma}
- \sum_{i\sigma} (\mu -\frac{U_i}{2} ) n_{i\sigma} + \sum_{i} U_i
(n_{i\uparrow} - {1 \over
2} ) (n_{i\downarrow} - {1 \over 2})~. \label{E6}
\end{equation}
The trace over the fermion degrees of freedom cannot be performed
analytically due to the quadratic interaction of term $U$. To reduce the
problem to  a quadratic Hamiltonian we discretize the imaginary time 
$\beta=L \Delta \tau$  (Eq. \ref{E3}) and employ the Trotter approximation
\cite{Fye86,Fye87}
to decompose the full imaginary time evolution. The partition function 
can be expressed as,
\begin{equation}
Z=Tr {\rm e}^{-\beta \hat H} =Tr \left[ {\rm e}^{-\Delta \tau \hat H} 
\right]^L \approx  \left[ {\rm e}^{-\Delta \tau \hat K} {\rm e}^{-\Delta
\tau \hat P} 
\right]^L~, \label{E7}
\end{equation}
where $\hat K$ includes the quadratic part of $\hat H$ (Eq. \ref{E6})
\begin{equation}
\hat K = -t \sum_{ ij\sigma}
c_{i\sigma}^{\dagger} c_{j\sigma}
- \sum_{i\sigma} (\mu -\frac{U_i}{2} ) n_{i\sigma}~, \label{E8}
\end{equation}
while $\hat P$ the on-site interaction part 
\begin{equation}
\hat P = \sum_{i} U_i
(n_{i\uparrow} - {1 \over
2} ) (n_{i\downarrow} - {1 \over 2})~. \label{E9}
\end{equation} 
Then we apply the discrete Hubbard--Stratanovich transformation to
decouple the attractive interaction $U_i < 0$  \cite{Hus97,Hus98,Sca99}
\begin{equation}
{\rm e}^{+ \Delta \tau U_i (n_{i\uparrow} - {1 \over
2} ) (n_{i\downarrow} - {1 \over 2})} 
= \frac{1}{2} {\rm e}^{\frac{-|U|
\Delta \tau}{4}} \sum_{Si=\pm 1} {\rm e}^{ \lambda S_i (n_{i\uparrow} + 
n_{i\downarrow}-1)}~, \label{E10} 
\end{equation}
where
\begin{equation}
\cosh{( \Delta \tau \lambda)} = {\rm e} ^{\Delta \tau |U|/2}~. \label{E11}
\end{equation}

The replacement of the quartic interaction term between up and down
spin electron densities by an expression in which the densities couple
to a Hubbard-Stratonovich field makes all terms in the exponential
of Eq.~7 quadratic in the fermion operators.  The trace can then  be
performed analytically, leaving a product of determinants
which depend on the Hubbard-Stratonovich field.  Because the up and down
spin species couple to the field with the same sign, these two determinants
are identical, and hence their product is always positive.
So, unlike the case of the repulsive model where $n_{i\uparrow}-
n_{i\downarrow}$ couples to the field, there is no sign problem
in the attractive model, even in the presence of disorder in the interaction.

\begin{figure}[htb]
\centering
\vspace{1cm}
\resizebox{1.0\textwidth}{!}{%
\hspace{1.5cm}  \includegraphics{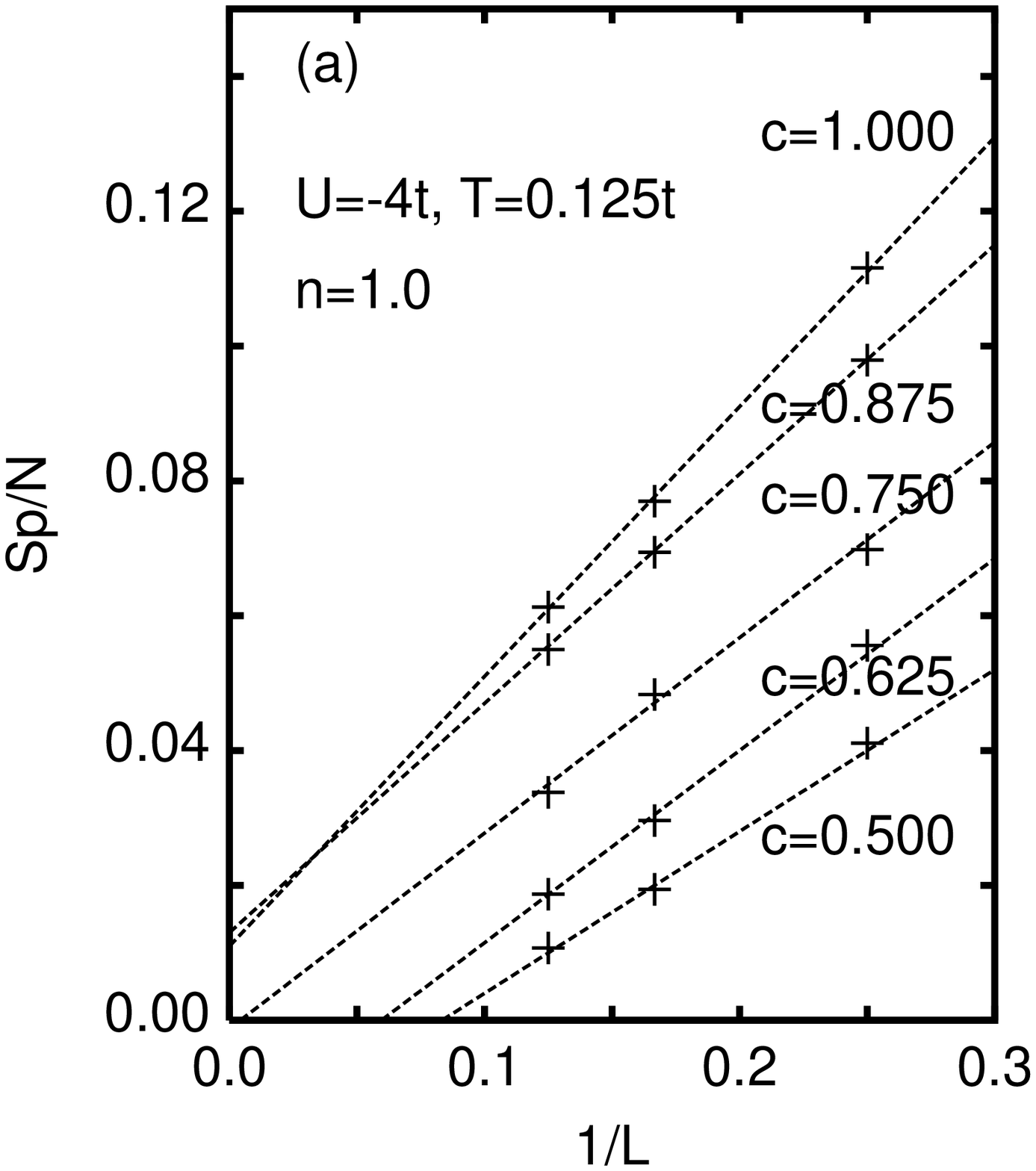} \includegraphics{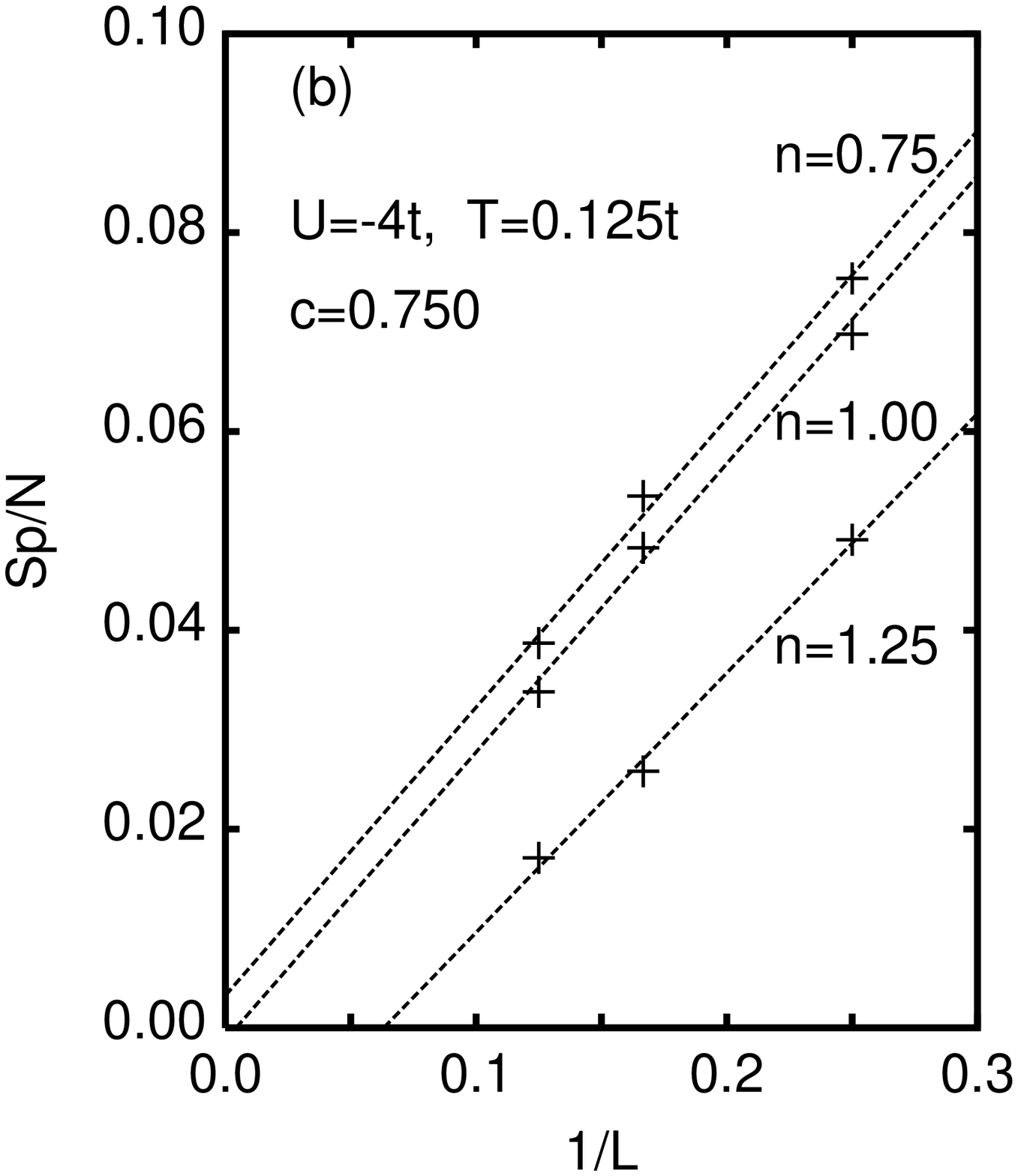}} 
\caption{ Finite lattice scaling of the structure factor $S_p$. Each
point represents the average for 20 different disordered lattice
configurations (for
$4\times 4$, $6 \times 6$ and $8 \times 8$) with relatively small error.
}
\end{figure}

\section{Percolation of Superconductivity}  

Using the above procedure we have simulated
the Hamiltonian (Eq. \ref{E6}) and calculated the  
pair structure factor $S_p$ (Eq. \ref{E5}) as a function of lattice
size $L$.  Spin-wave theory suggests that in
the ordered phase the finite size correction for
$S_p$ should be proportional to the inverse linear system size $1/L$\cite{Huse88}. 
Figure 2(a) presents the results of  finite size scaling of $S_p$ for 
various
concentrations of $U$ sites $c$.
The calculations were done for $U=-4t$ and $\beta=1/T=8/t$, and
the average  number of electrons per site was chosen to be $n=1$.
\begin{figure}[htb]
\centering
\resizebox{0.45\textwidth}{!}{ 
  \includegraphics{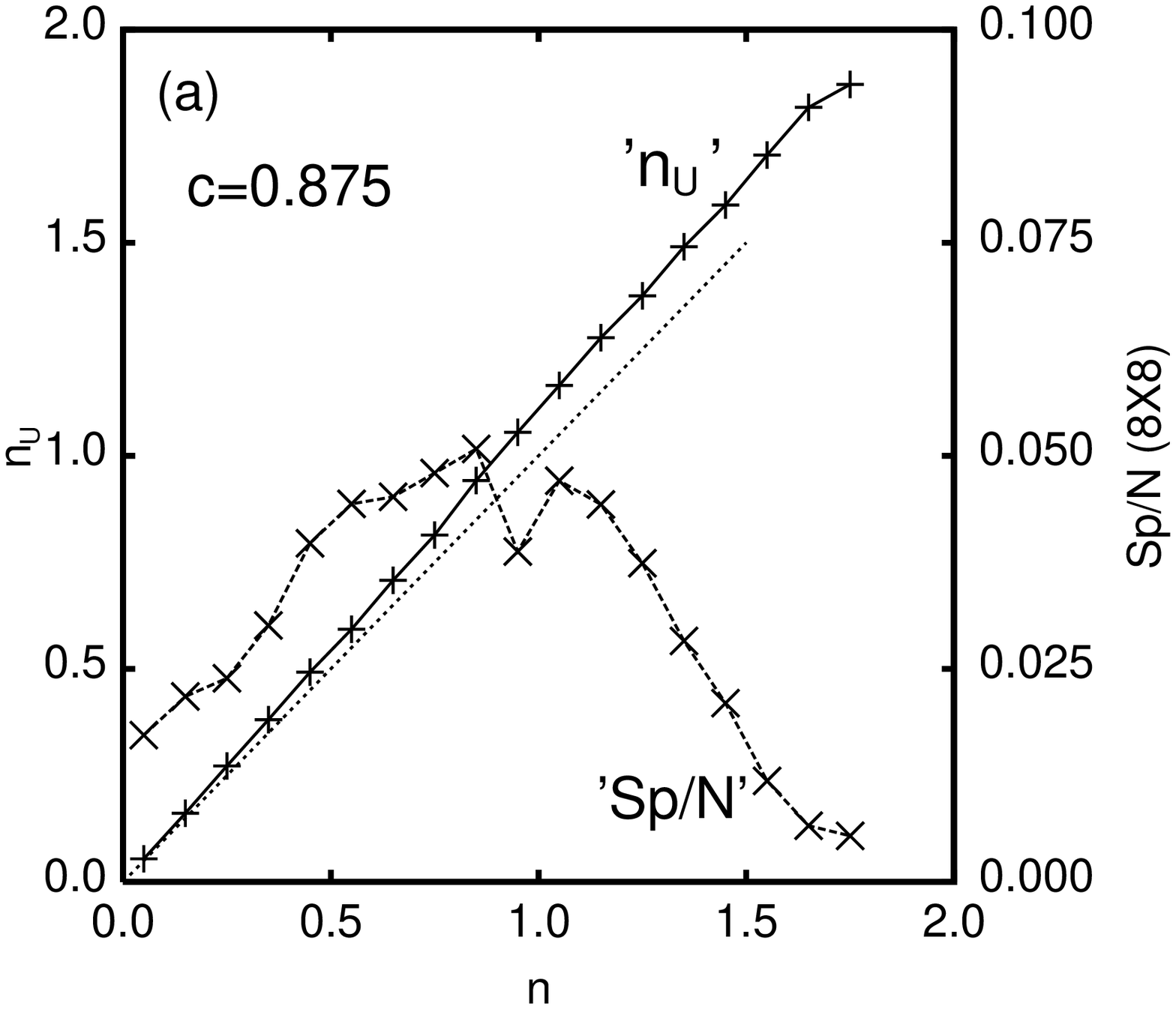}}
  
\resizebox{0.45\textwidth}{!}{ 
  \includegraphics{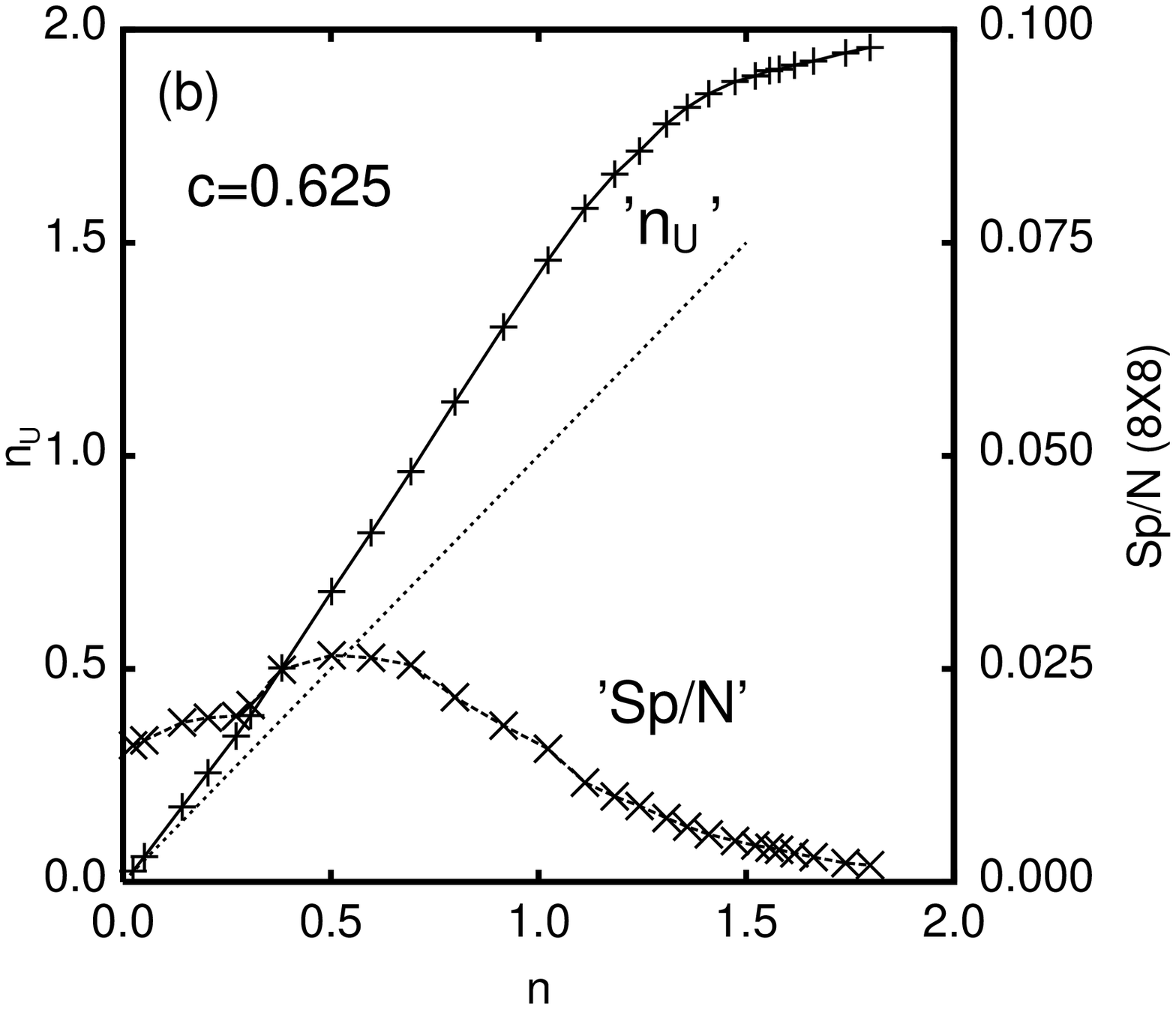}}
  
\resizebox{0.45\textwidth}{!}{ 
  \includegraphics{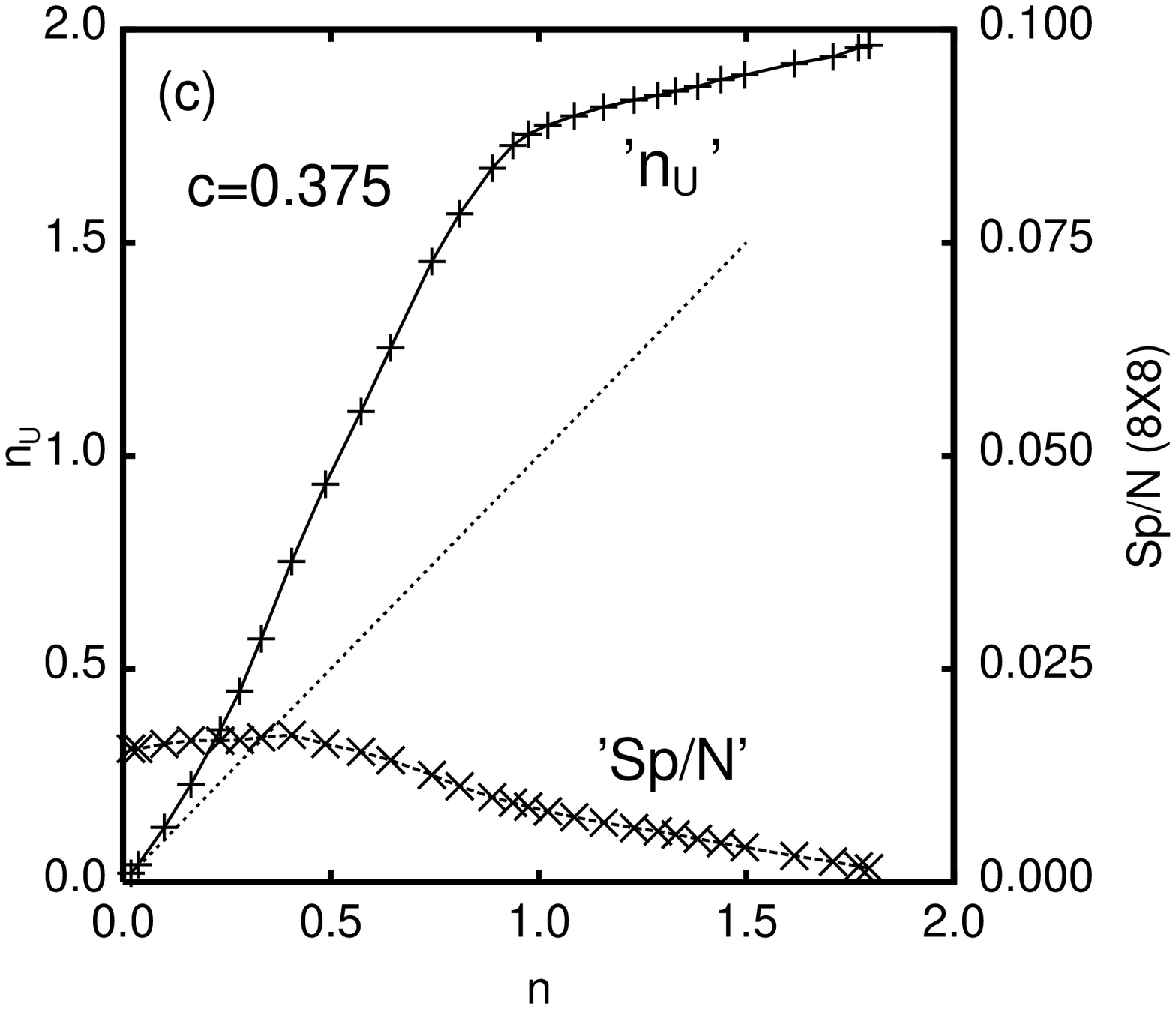}}
\caption{Charge on $U$ site $n_U$ and the structure factor of pair
correlation function $S_p$ versus system filling $n$ for a $8 \times 8$
lattice.} 
\end{figure}
One can easily see that   there exists  a critical concentration 
$c_0 \approx 0.75$. 

To see how $c_0$ depends on $n$ in  Fig. 2(b) the
concentration of $U$ centers is kept fixed at $c=0.750$, and we 
have plotted $S_p$ for various
system fillings
($n=0.75, 1.00$ and $1.25$). 
Here $n=1$ corresponds to the case at the  percolation threshold 
($c=c_0$). 
It is clear that for a larger average charge $n=1.25$
the system is not superconducting. A different situation emerges
for $n=0.75$ where $S_p(L \rightarrow \infty)$ appears to be slightly above
0, that is, the system is superconducting. 
Clearly,  disorder in the interaction $U$ has broken the particle--hole symmetry.
In fact, this is connected with the modification of the bare chemical potential
at $U$ sites, in the quadratic part of Hamiltonian  $\hat K$ (Eq. \ref{E6},\ref{E8}) 
\begin{equation}
\mu_{eff}= \left\{ \begin{array}{rl} \mu-U/2 & ~~~{\rm for}~~~ U_i=-|U|
\\ \mu  & ~~~{\rm for}~~~ U_i=0
\end{array}~. \right. \label{E12}
\end{equation}

In some previous work on the repulsive model\cite{Ulm98,Hus97,Hus98,Sca99}, 
disorder was inserted
in a way which preserved particle-hole symmetry by using
a separate shift in chemical potential depending on the  interaction
strength on each site.
This enabled simulations at lower temperature since it eliminated
the sign problem, but also was somewhat less physically motivated.

Interestingly, the structure factor associated with
the pairing correlation function $S_p(L
\rightarrow \infty)$ for $c=1$ is smaller than the  one for $c=0.875$.
This is due to the existence of a competing charge density
wave (CDW) phase at $n=1$.  A similar effect was found for the system 
with disordered site energies and the same attraction $U$ on 
every lattice site \cite{Hus97,Lit98}.    
Thus, in addition to phase fluctuation
of the pairing potential\cite{Gho98,Tri01}, it is clear that
the distribution of charges $n_i$ on the lattice 
(charge fluctuations) is also important for
the destruction of superconductivity.
Indeed, previous work suggests that percolation of the 
occupied sites is a crucial phenomenon in this process\cite{Lit00}.

Examining this effect further, one can
look for  the change of the  pairing factor correlation
function $S_p$ and simultaneously the average occupation of $U$ sites 
($n_U$) (where $U=-|U|$). Figures 3a-c show these
 quantities as a function of average site occupation $n$ for a finite system
$8\times8$ for $c=0.875$ (a), $c=0.625$ (b) 
and c=0.375 (c), respectively. Note that $S_p$
does not display particle-hole symmetry. 
The dotted lines in Fig. 3(a-c) represent the occupation of attractive 
sites
for a simplified Hamiltonian, where $\hat K$ reads
\begin{equation}
\hat K = -t \sum_{ ij\sigma}
c_{i\sigma}^{\dagger} c_{j\sigma}
- \sum_{i\sigma} \mu n_{i\sigma}~. \label{E13}
\end{equation}
It is easy to see   that in case of the full Hamiltonian
(Eqs. \ref{E2},\ref{E6}) the
system displays charge fluctuation ($n_U \neq n$) and comparing $n_U$
curves to the dotted lines
one can see that charge fluctuations are  stronger for smaller $c$. 
In our case, for large enough interaction $U$, the small concentration  
of $U$ centers  leads to  double occupation in almost all attractive sites.
This makes it harder for electrons
to hop to another $U \neq 0$ site. Nevertheless the system can be
metallic because   the other electrons which occupy 
the $U=0$ site can easily hop 
to the other $U=0$  sites. This result is consistent with the results
obtained by Litak and Gy\"{o}rffy \cite{Lit00}. The small dip visible in
the pairing structure factor $S_p$
around $n=1$ (in Fig. 3a) can be  associated with the creation of 
a CDW order. This effect  is not present in 
Figs. 3b-c as CDW expected to disappear
faster than superconductivity \cite{Hus97,Lit98}. 

\section{Conclusions}
 
We have examined the problem of percolating superconductivity in the  
context of a random $U$
Hubbard model. We have studied the case where $U_i$ is $-|U|$  
and 0 with probability $c$ and $1-c$ respectively by the quantum Monte
Carlo method.  For the half--filled system $n=1$ we obtain
the critical concentration $c_0=0.75$. 
As we expected the transition is due to phase
and charge fluctuations and depends strongly on the density  $n$.
In that sense our results agree qualitatively with those obtained
by Litak and Gy\"{o}rffy\cite{Lit00}. 
Moreover, the present approach is exact as opposed to
earlier work\cite{Lit00}
employing the mean field approximation.
We find that the onset of superconductivity in
this diluted, attractive Hubbard model is linked to 
percolation of the order parameter and charge fluctuations.   

\section*{Acknowledgements}
GL acknowledges a partial support by the Polish State Committee for
Scientific Research (KBN), Project No. 5 P03B 00221. 
The work of RTS was supported by NSF--DMR--9985978. TP snd RRdS are 
grateful to Brazilian agencies FAPERJ and CNPq for financial support.


\begin{thebibliography}{99}

\bibitem{Mic97} R. Micnas, S. Robaszkiewicz, in {\it High-T$_c$
Superconductivity 1996: Ten Years after the Discovery, NATO ASI Appl.
Sci.} E {\bf 343} (1997) 31.

\bibitem{Eme95} V.J. Emery, S.A. Kivelson, {\it Nature}  {\bf 374} (1995) 434.

\bibitem{Lin90}
V.J. Emery, S.H. Kivelson, and H.Q. Lin,
{\it Phys. Rev. Lett.} {\bf 64} (1990) 475.

\bibitem{Scalapino01}
D.J. Scalapino and S.R. White, 
{\it Foundations of Physics} {\bf 31} (2001) 27, and references cited therein.

\bibitem{Dagotto94}
E. Dagotto, {\it Rev. Mod. Phys.} {\bf 66} (1994) 763, and references cited therein.

\bibitem{Wil89} J.A. Wilson, {\it Int. J. Mod. Phys.} {\bf B 3} 691 (1989).

\bibitem{Mic90} R. Micnas, J. Ranninger and S. Robaszkiewicz, {\it Rev.
Mod.  Phys.} {\bf 62} (1990) 113.

\bibitem{Wil00} J.A. Wilson, {\it J. Phys: Condensed Matter} {\bf 12}
(2000) R517.

\bibitem{Tin92} C.S. Ting, in {\it Lattice Effects in High-T$_C$
Superconductors}, (World Scientific, Singapore 1992) 422.

\bibitem{Ale95}
A.S. Alexandrov and N.F. Mott, {\it Polarons and Bipolarons}
(World Scientific, Singapore 1995).

\bibitem{Ale99}
A.S. Alexandrov, and V.V. Kabanov, {\it Phys. Rev.} {\bf B} 59, (1999) 13628.

\bibitem{Ran95}
J. Ranninger, J.M. Robin and M. Eschrig, {\it Phys. Rev. Lett.} {\bf 74}
(1995) 4027.

\bibitem{Ran98}
J. Ranninger and A. Romano, {\it Phys. Rev. Lett.} {\bf 80}  (1998) 5643. 

\bibitem{Tri98}
N. Trivedi and M. Randeria, {\it Phys. Rev. Lett.} {\bf 75} (1998) 315. 

\bibitem{Levi99}
Y. Levi,  I. Felner, U. Asaf, and O. Millo,
{\it Phys. Rev.} {\bf B 60} (1999) R15059.

\bibitem{Lit00}  G. Litak and B.L. Gy\"{o}rffy, {\it Phys. Rev.} {\bf B 62}
(2000) 6629.

\bibitem{Ulm98} M.Ulmke, P.J.H. Denteneer, V. Janis, R.T. Scalettar, A.
Singh, D. Vollhardt, and G.T. Zimanyi, {\it Europhysics Lett} {\bf 42}
(1998) 655.

\bibitem{Gho98} A. Ghosal, M. Randeria, and N. Trivedi,
       {\it Phys. Rev. Lett.} {\bf 81} (1998) 3940.

\bibitem{Tri01} N. Trivedi, A. Ghosal, and M. Randeria,   
       {\it Int. J. Mod. Phys.}{\bf B 15} (2001) 1347.

\bibitem{Gyo91} B.L. Gy\"{o}rffy, J.B. Stauton, G.M. Stocks, {\it Phys.
Rev.} {\bf B 44} (1991) 5190.

\bibitem{Bla81} R. Blankenbecler, D.J. Scalapino, and R.L. Sugar,
 {\it Phys. Rev.} {\bf D} 24, (1981) 2278.

\bibitem{Hus97} C. Huscroft and R.T. Scatettar, {\it Phys. Rev.} {\bf B 55}
(1997) 1185.

\bibitem{Hus98} C. Huscroft and R.T. Scalettar, {\it Phys. Rev. Lett.}
{\bf 81} (1998) 2775.

\bibitem{Sca99} R.T. Scalettar, N. Trivedi, and
C. Huscroft, {\it Phys. Rev.} {\bf B 59} (1999) 4364.

\bibitem{Fye86} R.M. Fye, {\it Phys. Rev.} {\bf B  33} (1986) 6271.

\bibitem{Fye87} R.M. Fye and R.T. Scalettar, {\it Phys. Rev.} {\bf B 36} (1987) 3833.   
\bibitem{Huse88} D. A. Huse, {\it Phys. Rev.} {\bf B 37}, 2380 (1988).

\bibitem{Lit98} G. Litak, B.L. Gy\"{o}rffy and K.I. Wysoki\'nski,
 {\it Physica} {\bf C 308} (1998) 132.

\end{thebibliography}
\end{document}